\documentstyle[12pt,epsfig]{article}
\def\beq{\begin{equation}}   \def\eeq{\end{equation}}
 
\newcommand{\gsim}{\lower.7ex\hbox{$
\;\stackrel{\textstyle>}{\sim}\;$}}
\newcommand{\lsim}{\lower.7ex\hbox{$
\;\stackrel{\textstyle<}{\sim}\;$}}

\newcommand{\ra}{\rightarrow}

\begin{document}

\begin{titlepage}
\renewcommand{\thefootnote}{\fnsymbol{footnote}}

\begin{flushright}
TPI-MINN-13/97-T\\
UMN-TH-1542/97
\end{flushright}

\vspace{0.5cm}

\begin{center}
\baselineskip25pt

{\large\sc An Illustrative Example of How Quark-Hadron Duality
Might Work
          }  

\end{center}

\vspace{0.5cm}

\begin{center}
\baselineskip12pt

\def\thefootnote{\fnsymbol{footnote}}

{\sc B. Blok$^{1}$, M. Shifman, $^{2}$ and Da-Xin Zhang$^{1}$}

\vspace{0.5cm}

$^1$ Department of Physics, Israel Institute of Technology,
Technion, Haifa 32000, Israel

$^2$  Theoretical Physics Institute, University of Minnesota, 
Minneapolis,  MN 54555, USA

\end{center}

\vspace{1cm}

\begin{abstract}

We discuss  the issue of the local quark-hadron duality 
at high energies in two-  and four-dimensional
QCD. A mechanism of the dynamical realization of the 
quark-hadron duality in two-dimensional QCD in the limit of  large 
number of 
colors, $N_c\rightarrow\infty$, (the 't Hooft model) is considered. 
 A similar mechanism of 
dynamical smearing may be relevant in 
four-dimensional QCD. Although particular details of our results are 
model-dependent,  the general features
 of the duality implementation conjectured previously get further 
support.

\end{abstract}

\end{titlepage}

\section{Introduction} 

 In the recent years the focus of applications of the operator product
expansion (OPE) \cite{Wils1} has shifted towards the processes with 
the 
essentially Minkowskean kinematics. Perhaps, the most well-known 
example
is the theory of inclusive decays of heavy flavors (for a review 
see e.g.
Ref. \cite{Shif1}). This fact, as well as the increased demand 
for more accurate 
predictions, puts forward the study of the 
quark-hadron 
duality as an urgent task.

A detailed definition of the procedure which goes under the 
name
of the quark-hadron duality (a key element of every calculation 
referring to 
Minkowskean quantities) was given in Refs. \cite{Shif1,Chib1}. In a 
nut shell,
a {\em truncated} OPE is analytically continued, term by term, from 
the 
Euclidean to the Minkowski domain \footnote{Moreover, usually one 
deals 
with the practical version \cite{NSVZ} of OPE, see Ref. \cite{Chib1} for 
further 
details.}. A 
smooth quark curve obtained in this way is
supposed to coincide at high energies (energy releases) with the 
actual
hadronic cross section. 

If duality is formulated in this way, it is perfectly obvious that
at finite energies 
deviations from duality must exist. The difference between the 
measured physical cross section and a smooth OPE prediction
will be referred to as an oscillating/exponential (duality violating) 
component.
In Ref. \cite{Shif1} it was shown
that if we knew the leading asymptotic behavior of the high 
order
terms in the power series we could evaluate this component. 
Unfortunately, very little is known 
about this
aspect of OPE, and we have to approach the problem from the other 
side --
either by modeling the phenomenon \cite{Chib1} or by studying 
some
general features of the appropriate spectral densities. One can also 
try to 
approach the problem purely phenomenologically (for recent 
attempts
see e.g. \cite{Blok1}).

An illustrative spectral density, quite instructive in the studies of the
issue of the quark-hadron duality,  was suggested in \cite{Shif1},
\beq
{\rm Im}\, \Pi = \mbox{Const.} \, \frac{N_c}{2} \, \sum_{n=1}^\infty 
\delta 
({\cal E} - n)
\label{modim}
\eeq
where
$$
{\cal E} = \frac{s}{\Lambda^2}\, ,
$$ 
and from now on we will drop an inessential constant in front of the 
sum.
The color factor $N_c$ is singled out for convenience.
The imaginary part above represents, for positive values of $s$, a 
sum of 
infinitely narrow
equidistant resonances, with equal residues. The distance between 
the resonances is $\Lambda^2$. 
It defines $\Pi (q^2)$ 
everywhere
in the complex plane $q^2$, through the standard dispersion relation, 
up to an 
additive constant which can be adjusted arbitrarily. It is not difficult 
to see
that the corresponding correlation function
\beq
\Pi (q^2) = -\frac{N_c}{2\pi} \left[ \psi (\varepsilon ) + 
\frac{1}{\varepsilon}\right]
\label{MSmodel}
\eeq
where $\psi$ is the logarithmic derivative of Euler's gamma function, 
and
$$
\varepsilon = -\frac{q^2}{\Lambda^2} = - {\cal E} \, .
$$
In the Minkowski domain ${\cal E}$ is positive, in the Euclidean 
domain
$\varepsilon$ is positive. Then, the asymptotic expansion of $\Pi 
(q^2)$
in deep Euclidean domain is well known,
\beq
\Pi (q^2) \ra -\frac{N_c}{2\pi} \, \left[ 
\ln \varepsilon +\frac{1}{2\varepsilon} - \sum_{n=1} \left( 
-1\right)^{n-1}
\frac{B_{n}\varepsilon^{-2n}}{2n}\right]
\eeq
where $B_{n}$ are the Bernoulli numbers. At large $n$ they  grow 
factorially, 
as
$B_{n}\sim (2n)!$ (see \cite{Wang}, page 23). We deal with the  sign 
alternating series.

Although the spectral density (\ref{modim}) is admittedly a model, it 
was 
argued \cite{Shif1} that a similar factorial growth of the coefficients 
in the 
power (condensate) series is a general feature.

The spectral density (\ref{modim}) may be relevant in the limit
of the large number of colors, $N_c\ra\infty$, when all mesons are 
infinitely
narrow. This limit is not realistic, however. Moreover, in this 
limit
the {\em local} quark-hadron duality, as we defined it, {\em never} 
takes place
since even at high energies the hadronic spectral density never 
becomes 
smooth, even approximately. One can smear it by hand, of course, but 
then deviations 
from the local duality
will be determined not only by the intrinsic hadronic dynamics, as is 
the case 
in the real world, but also by particular smearing procedure
-- they will depend on  
the weight function chosen for smearing, the interval of smearing 
and so on. 
In the actual world the smearing occurs {\em dynamically}, since at 
high 
energies
the resonance widths become non-negligible. The limits of 
$E\ra\infty$
and $N_c\ra\infty$ are not interchangeable. 

Here we suggest and  study more realistic (dynamically smeared) 
spectral 
densities
compatible with all general properties of Quantum Chromodynamics.
Starting from  infinitely narrow resonances, as in Eq. (\ref{modim}),  
we   
introduce finite widths, ensuring smooth behavior.
Technically, in the first part of the paper the problem of duality is 
analyzed in
the two-dimensional 't Hooft model \cite{tHoo1} (see also
\cite{LTLY} -- \cite{Bars}). The quark confinement in this model is
built-in. We then try to abstract general features of this solution,
which may persist in QCD. In the second part of the paper an attempt 
is made to work out the 
same mechanism  in four dimensions.

Dynamical ``smearing" of the spectral densities taking part in QCD 
due 
to nonvanishing resonance widths, exhibits the same  features of the
high-energy behavior as was suggested in Refs. [2] and [3] on the 
basis of
rather naive  models, e.g. instanton models. Namely, the approach of 
the spectral density to 
the smooth limit (the deviation from duality) is exponential, with 
oscillations.
This pattern seems to be general and may be considered now 
as a 
well-established model-independent fact. At the same time, the 
exponent
 determining the rate of the damping of the duality violations 
depends on 
details of the large distance dynamics. We were unable to find 
it from  first principles, and had to settle for model-dependent 
determinations.  

Let us note that several  useful results on the relation between 
duality 
violations and the divergence of the power expansion were 
presented in Ref. 
\cite{Zhit1}, and we incorporate them
\footnote{We strongly disagree, however, with some particular 
calculations 
and expressions,
presented in this paper.}. 
 The very idea of  using the 't 
Hooft model
as a theoretical laboratory adequate to the problem was formulated 
there. 
Moreover, it was noted \cite{Zhit1} that the leading asymptotics of 
the 
high-order coefficients does not require the knowledge of the exact 
mass 
spectrum.
Suffice it to know the leading term in the $n$ dependence, where $n$ 
is the 
radial excitation number, which immediately translates in the 
leading factorial 
behavior of the coefficients. In particular,
the position of the low-lying states is irrelevant  in the regime with 
the 
factorially growing coefficients. This is a specific feature of the
asymptotic series.

The $\psi$ function model was originally suggested in Ref.
\cite{Shif1} for the heavy-light quark systems. In Ref. \cite{Zhit1} it 
was 
noted that it was more appropriate for the light-light quark systems
since in the heavy-light  systems the resonances are not expected
to be equidistant. Straightforward quasiclassical estimates
yield in this case that the meson energies (measured from the heavy 
quark mass) asymptotically scale as $\sqrt{n}$. (See also 
 Ref. \cite{CDN}, where this scaling law is reproduced in a linear  
potential  model.)
In the present paper we further develop the $\psi$ function {\em 
ansatz}
adapting it for the light-quark systems. 

The paper is organized as follows. Section 2 briefly reminds the basic
strategy of the OPE-based calculations. Here we outline the main 
elements of the analysis to be opresented below. 
In Sect. 3  we first discuss 
the 
't Hooft model in the leading $1/N_c$ approximation. Then we 
calculate the resonance
widths and evaluate  the 
polarization operator of the two scalar currents in   taking 
into 
account 1/$N_c$ corrections.  In  Sect. 4 we
present  qualitative arguments why the implementation
 of duality obtained in the 't Hooft model must be also relevant,
at a qualitative level, in four-dimensional QCD.
Sect. 5 gives  conclusions and outlines possible physical 
applications 
of our results.

\section{OPE-Based Strategy, Sum Rules and Duality}

The issue of ``deviations from duality" caused much confusion in the 
recent literature. Therefore, to begin with, we remind 
 the basic strategy revealing, in clear terms, where
the deviations can occur  and where there can be no deviations.

First of all, all calculations based on the operator product expansion 
are carried out in the Euclidean domain. Only away from the
physical cuts this procedure is well-defined. Any calculation consists 
of several crucial elements: identification of the operators which can 
appear in the expansion, separation of hard virtual momenta (higher 
than a normalization point $\mu$) from soft virtual momenta (lower 
than $\mu$) and, finally, calculation of the expansion coefficients in 
front of
relevant operators. The latter is carried out in terms of quarks
and gluons. That is why the normalization point $\mu$ must be 
chosen sufficiently high, and the calculation must be done in the 
Euclidean domain. Even if in some problems calculations are 
conveniently presented in such a way as if they were done in  the 
Minkowski domain, actually the corresponding results must be 
understood as an analytic continuation. 

The connection between the Euclidean predictions and measurable 
quantities is established {\em via} dispersion relations. 
In this way one can get certain sum rules. A large variety of them is 
offered on the market. If the Euclidean quantity is appropriately 
chosen, the (Euclidean) operator product expansion {\em converges}.
The best-known example of such an appropriate choice is provided 
by the SVZ sum rules \cite{SVZ} obtained by virtue  of the Borel 
transformation of the dispersion representation. Consider, for 
instance, the model of Ref. \cite{Shif1}, see Eq. (\ref{MSmodel}).
The $1/Q^2$ expansion of $\Pi (q^2 )$ in the Euclidean domain is 
factorially divergent. At the same time, the $1/M^2$ expansion of the 
Borel-transformed quantity has a finite radius of convergence.
Indeed, the SVZ sum rule in the case at hand has the form
\beq
\frac{N_c\Lambda^2}{2\pi}\, \frac{{\rm e}^{-\Lambda^2/M^2}}{1-
{\rm e}^{-\Lambda^2/M^2}} =\frac{1}{\pi}\int {\rm e}^{-s/M^2}
{\rm Im}\, \Pi (s) ds\, ,
\label{SR}
\eeq
(in the right-hand side $\Lambda^2$ plays the role of the  inverse 
slope of the 
Regge trajectory).
Assume that the left-hand side was calculated theoretically, using 
OPE, as an expansion in $1/M^2$. The domain of convergence of the
power 
expansion is determined by the position of the nearest singularity
in the complex $M^2$ plane. It is quite obvious
that the expansion converges at
\beq
|M^2| > \frac{\Lambda^2}{2\pi}\, .
\label{RC}
\eeq
Not only the radius of convergence is finite, due to the factor
$2\pi$ in the denominator the domain of convergence extends to 
quite low values of $M^2$. If $\Lambda^2 \sim $ 1 to 2 GeV$^2$,
the power series is convergent at as low values of $M^2$ as
0.4 GeV$^2$. This fact was empirically observed long ago \cite{SVZ}.

If we consider the SVZ sum rule inside the convergence domain of 
OPE (see Eq. (\ref{RC}), 
and  are aimed at predicting the exponential integral
on the right-hand side of Eq. (\ref{SR}) {\em per se},
it is meaningless to speak about deviations from OPE.
In this formulation of the problem {\em there are no deviations}. 

The problem of deviations arises when we try to predict the spectral 
density
Im $\Pi$ point-by point, at large $s$, or   certain integrals
of Im $\Pi$, not directly reducible to ``good" sum rules
 of the type (\ref{SR}). Certainly, if we assume that Im $\Pi$ is 
smooth starting from some boundary value $s_0$, Eq. (\ref{SR})
allows us to predict Im $\Pi (s)$ at $s>s_0$ unambiguously. 
If an oscillating component is allowed,  however,  one 
can always invent such a wild oscillating function,
which, being integrated with the exponential weight,
gives a contribution on the left-hand side of the sum rule
(\ref{SR}) less than the last term of the power expansion retained, no 
matter how small 
this last term is. This component is referred to as duality-violating. 
Clearly, Quantum Chromodynamics admits only
very specific oscillating/exponential components in the spectral 
densities, if at all. 
The question is what particular oscillating/exponential functions are 
allowed by QCD dynamics. We first try to answer this question in a 
simpler context of the 't Hooft model, and then pass to discussion of 
QCD. 

\section{The 't Hooft model}

The aim of this section is to study the quark-hadron duality in 
the 
't Hooft model. We shall first briefly review the 't Hooft model in the 
$N_c
\rightarrow \infty$ limit and discuss  $1/N_c$ corrections in
this 
model. Our  original contribution is  calculating the resonance widths 
for high excitations.
 Then, using these widths, we 
suggest an {\em ansatz} for  
 the asymptotic behavior of  the polarization operator of two 
scalar currents, based on the Breit-Wigner approximation.
 This resonance-saturated polarization operator
will be referred to as phenomenological. We will confront it with
the truncated power expansion. The difference between these two 
expressions gives an idea of the  duality violation.

\subsection{ The  't Hooft model: generalities}

 Two-dimensional  QCD is described by the  
Lagrangian
\begin{equation}
L=- \frac{1}{4}G^a_{\mu\nu}G^a_{\mu\nu}+\bar q^{f}(i\not\!\!{D}  - 
m_f)q^f\, ,
\label{lag}
\end{equation}
where $f$ is the flavor index. Since the multiflavor aspect is 
irrelevant for our problem, we  shall consider, for simplicity, one 
flavor; correspondingly, the index $f$ will be omitted hereafter.
If the gauge coupling of the theory is $g$, it is convenient to 
introduce an effective coupling,
\begin{equation}
\bar g^2 = g^2 N_c\, ,
\end{equation}
which stays constant in the limit $N_c\rightarrow\infty$. 
The coupling constant $g$ has dimension of mass; 
it sets the scale for all dimensional quantities in the chiral limit, i.e.
$m\rightarrow 0$. We introduce the scale
\begin{equation}
\mu^2 = \frac{\bar g^2}{\pi}=\frac{g^2N_c}{\pi}\, ,
\label{scale}
\end{equation}
and  measure all quantities in these units, e.g.  the quark mass
$$
\gamma^2 = m^2 /\mu^2\, ,
$$
while  the mass of the $n$-th meson
$$
\mu_n^2 = m_n^2/\mu^2\, ,
$$
and so on \footnote{Normalization of the coupling constant $g$ 
coincides 
with that 
adopted  in Refs. \cite{tHoo1} and  \cite{CCG}, but differs from 
that in Ref. \cite{Bars}.}.  

As was shown by 't Hooft,  the model with the Lagrangian
(\ref{lag}) is exactly solvable in the 
limit
$N_c\rightarrow\infty$. The bound state spectrum includes an 
infinite 
number of bound states whose masses lie on an almost linear 
trajectory.
The properties of these bound states are described by the 't Hooft 
equation, 
\begin{equation}
\mu^2_n\phi_n(x)=\frac{(\gamma^2 -1)\phi_n (x)}{x(1-x)}-
-\!\!\!\!\!\int^1_0\frac{\phi_n(y)dy}{(x-
y)^2}.
\label{equation}
\end{equation}
Here the integral is understood as a ``principal value",
$$
\frac{1}{p^2} = \lim_{\epsilon\rightarrow 0}
\frac{1}{2}\left[ \frac{1}{(p+i\epsilon )^2} +\frac{1}{(p-i\epsilon 
)^2}
\right] \, .
$$
Moreover, $x$ is the momentum fraction of the meson carried by
the antiquark (in the infinite momentum frame), while $1-x$ is that
of the quark; 
$\phi_n(x)$ is the wave function of the $n$-th bound state.

The integral equation (\ref{equation}) must be solved with the 
boundary
conditions:
\begin{equation}
\phi_n (x)
\rightarrow  \left\{
\begin{array}{c}
x^\beta \, ,\,\,\, x\rightarrow 0\\
         (1-x)^\beta\, ,\,\,\, x\rightarrow 1
\end{array}\right.  .
\label{boundary}
\end{equation}
Here $\beta$ is the smallest (in the absolute value) root of the 
equation
\beq
\pi\beta{\rm cot} (\pi\beta ) =1-\gamma^2.
\label{b2}
\eeq

Below we shall be interested in the massless case, $\gamma =0$.
In the massless limit, the lowest-lying state (the ``pion") can be 
found from Eq. (\ref{equation}) analytically. Indeed, the solution
$\phi_0 =$ Const., corresponding to $\mu_0 = 0$, obviously goes 
through. We are interested, however, in  highly excited states, 
$n\gg 1$. 

In the original paper \cite{tHoo1} 't Hooft suggested the 
following approximation for $\phi_n(x)$
at large $n$, $x$ not too close to 0 and 1:
\beq
\phi_n(x)=\sqrt{2} \sin {(n\pi x)}.
\label{wf2}
\eeq
  Recent calculations 
\cite{KS}, exploiting a new improved numerical procedure (the so 
called spline 
method), 
show that for the massless case a better approximation for large $n$ 
is 
\begin{equation}
\phi_n(x)=\sqrt{2}{\rm cos} (\pi n x)\, .
\label{wf1} 
\end{equation}
This formula works very well numerically everywhere except the 
very endpoints
$ x=0,1$, and a slight 
$x$-dependent shift when $x \sim 0.5$.

Note that the wave functions (\ref{wf1}) satisfy the proper boundary 
conditions 
for 
the 
massless case, which corresponds to  $\beta$ equal to zero, and, 
hence,
\beq
\phi_n (0)=C, \phi_n (1)=PC.
\label{bq3}
\eeq
Here $P=1$ for the states with the even parity and $P=-1$ for the 
states 
with the
odd parity, and the constant $C\ne 0$.

The mass spectrum in the massless case was found by 't Hooft. The 
asymptotic behavior of
$\mu_n^2$  is 
\begin{equation}
\mu^2_n= \pi^2 n \left[ 1+O(\ln (n)/n + ... \right]\, .
\label{masses}
\end{equation}

Note  that the mass formula (\ref{masses}) 
does not depend on the choice of the wave function, Eq. (\ref{wf1}) 
or the 
't Hooft choice,
at least in the leading in $n$ approximation \cite{KS}.

\subsection{The meson widths}

 As was already mentioned, in the limit 
$N_c\rightarrow 
\infty$ the bound states in the 't Hooft model are stable, their 
widths vanish. However, once one takes into account the leading
$1/N_c$ correction, the resonances begin to decay. In the first 
order in $1/N_c$ expansion there are only two-particle decays
$a\rightarrow b+c$. The relevant coupling constants $g_{abc}$ are 
given by the following formula
\cite{CCG,B,KS,E1,E2}:
\begin{equation}
g_{abc}=\mu^2\sqrt{\frac{\pi}{N_c}}
[1-(-1)^{(\sigma_a+ \sigma_b+ \sigma_c)}](f^+_{abc}+f^-_{abc})\, .
\label{decayconstant}
\end{equation}
Here $\sigma_a$ is the parity of the $a$-th resonance.
The constants $f^{\pm}_{abc}$ are determined from the following 
expressions:
$$
f^{\pm}_{abc}=\displaystyle\frac{1}{1-
\omega_{\pm}}\int^{\omega_{\pm}}_0
\phi_a(x)\phi_b(x/
\omega_\pm)\Phi_c\left( \frac{x-\omega_\pm}{1-\omega_\pm}
\right) 
$$
\begin{equation}
-\frac{1}{\omega_\pm}\int^1_{\omega_\pm}\phi_a(x)\Phi_b(x/
\omega_\pm)
\phi_c\left( \frac{x-\omega_\pm}{1-\omega_\pm}\right)
\, ,
\label{c}
\end{equation}
where $\omega_{\pm}$ are  two roots of the algebraic equation 
corresponding to the mass-shell condition,
\begin{equation}
m^2_a=\frac{m^2_b}{\omega} +\frac{m^2_c}{(1-\omega )} \, .
\label{massshell}
\end{equation}
The function $\Phi_a (x)$ is defined as
$$
\Phi_a (x)= -\!\!\!\!\!\int^1_0 dy \frac{\phi_a (y)}{(x-y)^2}.
$$

 Using the above expressions for the decay  couplings one can readily 
calculate the resonance 
widths 
in the leading $1/N_c$ approximation. They are given 
by
\begin{equation}
\Gamma_a=\frac{1}{8m_a}\sum_b\sum_c\frac{
g^2_{abc}
}{\sqrt{I(m_a,m_b,m_c)}}\, ,
\label{sum}
\end{equation}
where $I$ is the standard ``triangular" function,
\begin{equation}
I(m_a,m_b,m_c)=\frac{1}{4}[m_a^2-(m_b+m_c)^2][m_a^2-(m_b-
m_c)^2]\, .
\label{triangular}
\end{equation}
The sum in Eq. (\ref{sum}) runs over all mesons $b$ and $c$ with
the constraint $m_b+m_c <m_a$.

 Our task was to establish the asymptotic behavior of the 
widths, as a function of the excitation number, at large values of
$n$, in the leading $1/N_c$ approximation (all widths are
proportional to $1/N_c$; the excitation number $n$ will be 
temporarily called $a$ in this section). 
In order to find the widths we first computed analytically,  using the  
wave 
functions 
(\ref{wf1}), the overlap integrals (\ref{c}).  The answer can be 
expressed
via the integral sine and  cosine functions and was obtained
using the REDUCE program. Since it is very bulky it seems 
unreasonable to 
present here the final expression \cite{private}.
 After computing the overlap integral we 
performed numerically  summation over all possible $b$ and $c$
in Eq. (\ref{sum}) for $a$ up to  500. The result for the 
widths exhibits a remarkable pattern. The widths of the individual 
levels oscillate near a smooth square-root curve, see Fig. 1. This 
figure 
shows the width of the $a$-th state {\em vs.} $a$, up to $a= 500$.
 The result of averaging over the 
interval 
of 20 resonances is depicted in Fig. 2. We see that the curve of 
the
averaged resonance 
widths $\Gamma (a)\equiv \Gamma_a$ is very
well approximated by the function 
 \begin{equation}
\Gamma (a)=\frac{A\mu}{\pi^2 N_c}\sqrt{a}\left( 1+ {\cal O} 
(1/a)\right) \, ,
\label{widths}           
\end{equation}
where the parameter $\mu$ is
introduced in Eq. (\ref{scale}), and $A$ is a constant which will be 
given
 below.

\begin{figure}[htbp] \unitlength 1mm
\begin{center}
\begin{picture}(160,160)
\put(0,85){\epsfig{file=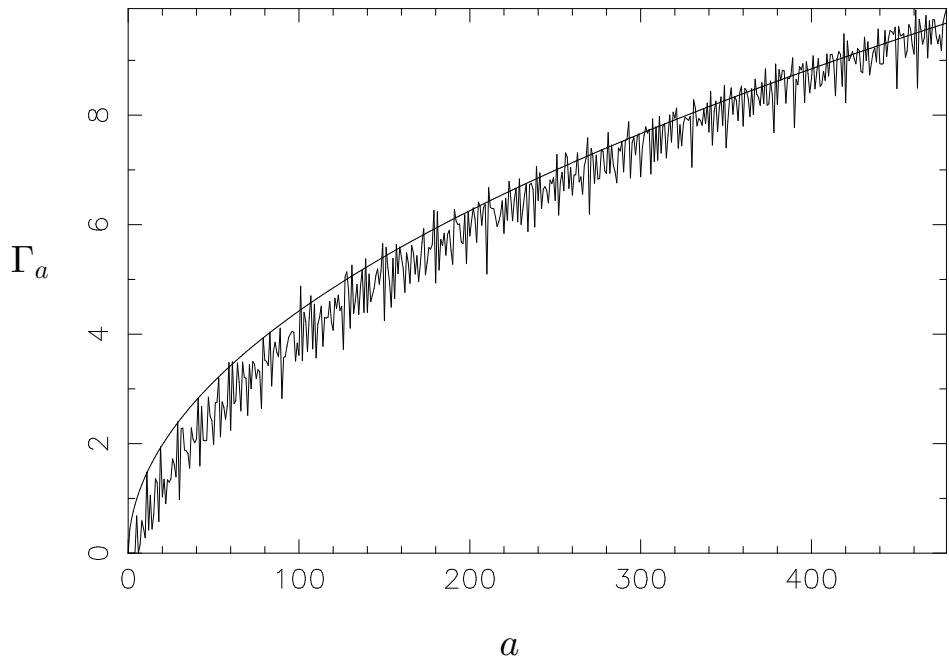,width=14cm,angle=270}}
\put(0,138){\large ${\Gamma_a}$}
\put(65,87){\large $a$}
\end{picture}
\caption{The width of the $a$-th state (in units
of $\mu \pi^{-2} N_c^{-1}$) as a function of $a$
up to $a=500$.}
\end{center}
\end{figure}

\begin{figure}[htbp] \unitlength 1mm
\begin{center}
\begin{picture}(160,160)
\put(0,85){\epsfig{file=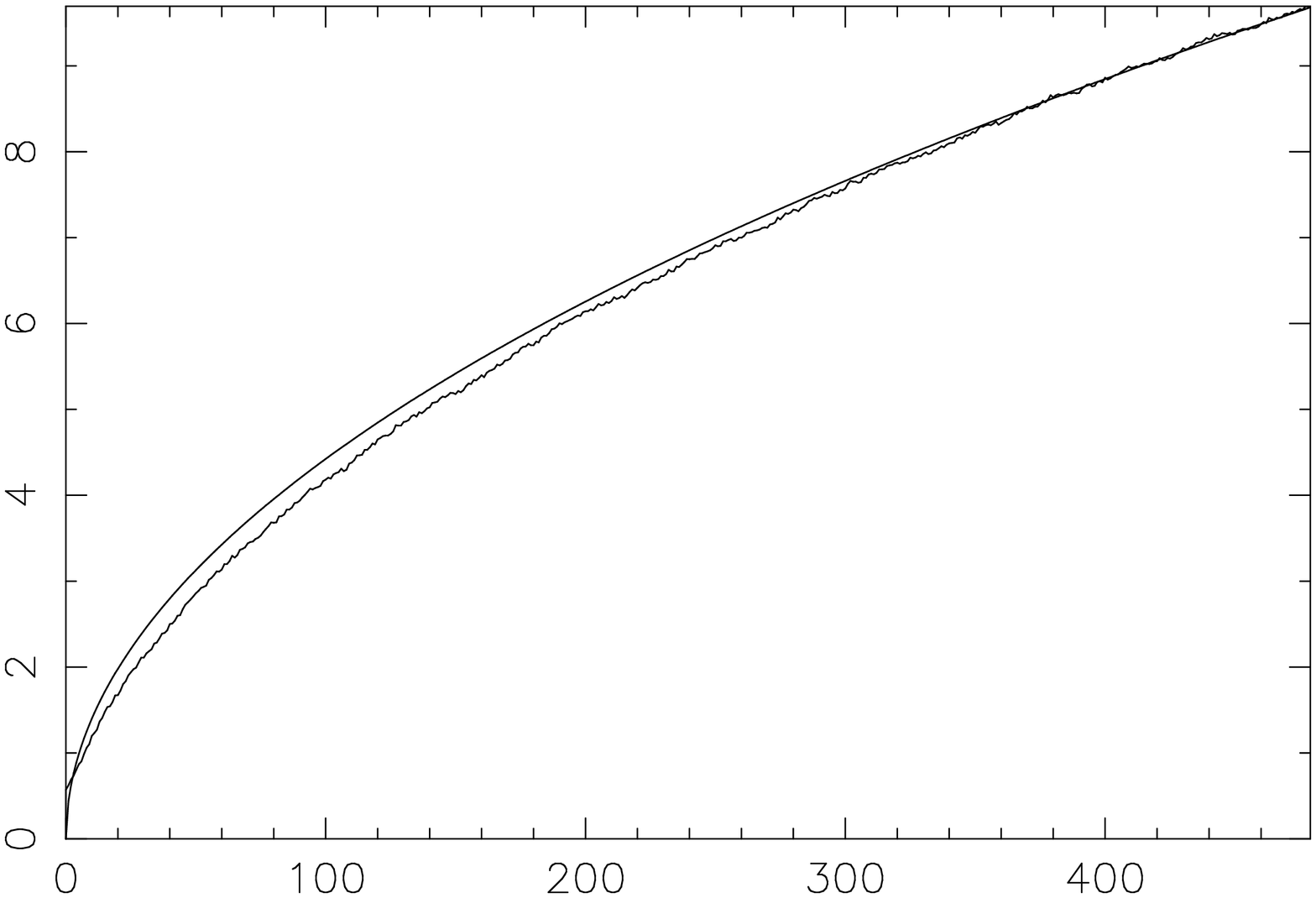,width=14cm,angle=270}}
\put(0,138){\large ${\Gamma_a}$}
\put(65,87){\large $a$}
\end{picture}
\caption{Smeared widths in the same interval of $a$ 
and the square-root fit, Eq. (21).}
\end{center}
\end{figure}

Since the square-root law (\ref{widths}) for the (averaged) widths is 
valid in such a large interval of the excitation numbers and 
turns out to be so accurate,  it seems  plausible that
this formula could be obtained analytically. This is an interesting 
question by itself, especially in  four-dimensional QCD.
Unfortunately, we were unable to find exact analytic solution so far.
Some qualitative arguments in favor of the exact square root 
dependence
are discussed in Sect. 4.
 The numerical value of the constant $A$   is 
\begin{equation}
A= 0.44\pm 0.05. 
\label{constant}
 \end{equation}
Below this result for the (averaged) widths will be used for
determining  the asymptotic behavior of the polarization operator
\footnote{Let us note that if  one  
calculates the widths with  the wave functions (\ref{wf2}), the 
square-root behavior of  
Eq. (\ref{widths}) is intact,  but  the value of the  constant $A$ is
different,  $A'\sim 0.007$, i.e.
$\sim 50$ times 
smaller. This fact indicates that the square-root law is not sensitive 
to the 
precise form of the wave functions, while the  value of the 
coefficient $A$ is.}.

Since the $1/N_c$ result for the decay width grows with $a$, one 
may worry about the $1/N_c^2$ corrections. 
If they grew with $a$ sufficiently fast this could invalidate 
Eq. (\ref{widths}) in the interval of the excitation numbers we are 
interested in, namely, $a=$ Const. $N_c$, where the constant above 
can be numerically large, but it does not scale with $N_c$. 
Using quasiclassical arguments (see Sect. 4) one can show that
the actual $1/N_c$ expansion parameter in Eq. (\ref{widths}) is
$\sqrt{a}/N_c$.  This means that at $a=$ Const. $N_c$
corrections $O(1/N_c^2)$ and higher are negligible. 

Concluding this section let us note that  the same square-root
was reported previously 
 in Ref. \cite{B}. We failed to reproduce the arguments of this work 
leading to the square-root law, however. What is 
important, the constant analogous to $A$ in Ref. \cite{B} is 
claimed to be proportional to $1/\sqrt{m}$ (!), and, 
thus, 
blows up for  massless quarks. This poses perplexing questions.
The coincidence looks completely accidental.

\subsection{The Breit-Wigner approximation. An Ansatz for the 
polarization
operator}

Once we had found the resonance widths, we can calculate the 
polarization operator in the Breit-Wigner approximation. Here we 
shall 
consider the most interesting case of the polarization operator of the 
two 
scalar currents. Let us start from this polarization operator in the 
$N_c\rightarrow\infty$ limit \cite{CCG,Z}.

Define the two-point function of the scalar currents $j =\bar q q$,
 \begin{equation}
\Pi (q^2)=i \int d^2x {\rm e}^{iqx}\langle
0\vert T\{j(x),j(0)\}\vert 0\rangle\, .
\label{Pol}
\end{equation}
In the $N_c\rightarrow\infty$ limit  $\Pi $ is given 
 by \cite{CCG}
\begin{equation}
\Pi(q^2)=
-\sum^{n=\infty}_{n=0}\frac{g^2_n}{q^2-m^2_n+i\epsilon}\, .
\label{pol1}
\end{equation}
Here the constants $g_n$ are the current residues
\begin{equation}
\langle 0\vert j(x) \vert n\rangle = g_n\, .
\label{residue}
\end{equation}   
Note that the residues vanish for even $n$ \cite {CCG}, so that the 
sum in Eq.
(\ref{pol1}) runs over odd $n$ only.
Below we shall be interested in the behavior of $\Pi (q^2)$ for large 
$|q^2| \gg \mu^2$. This behavior
is
 dominated by the terms in the sum (\ref{pol1}) with large $n\gg 1$
 \cite{CCG,Z,Zhit1}. 
In order to calculate this sum explicitly we then need the large 
$n$ behavior of $g_n$. It was determined in the 
same classical paper \cite{CCG} from the requirement of the 
compatibility 
of the expansion (\ref{pol1}) and the perturbation theory 
asymptotics
 in the $Q^2\rightarrow\infty$ limit,
\begin{equation}
\Pi (q^2)\rightarrow -\frac{N_c}{2\pi }{\rm ln}\frac{Q^2}{\mu^2}\, ,
\,\,\, Q^2\equiv -q^2\, .
\label{asymptotics}
\end{equation}
 The coefficients $g_n$ for sufficiently large odd $n$ must 
be 
independent of $n$ and are equal to
\beq 
g^2_n=N_c\pi\mu^2.
\label{lgn}
\eeq
 Then, taking into account the linear $n$ dependence 
of mass squared  one can approximate the polarization operator 
(\ref{Pol}) for sufficiently large $|q^2|$ by the $\psi$ function,
\begin{equation}
\Pi (q^2)-\Pi(0)= -\frac{N_c}{2\pi}
\psi(\sigma )\, ,\,\,  \, \sigma =\frac{Q^2}{2\pi^2\mu^2}+\frac{1}{2} \, 
.
\label{psi}
\end{equation}
We hasten to emphasize again that this formula is not supposed to 
work at non-asymptotic values of $Q^2$.  For instance, it does not 
contain the massless ``pion". Moreover, by shifting
a little bit the masses and residues of the low-lying resonances
we let $\Pi (q^2)$ ``breathe" at small $Q^2$ without changing the 
asymptotic behavior. 

\vspace{0.2cm}

What will happen if we take into account 
finite 
widths of the resonances? To answer this question we calculate $\Pi 
(Q^2)$ in the 
Breit-Wigner approximation.
(Continuation of the Breit-Wigner formula in the complex plane, 
away from the 
resonance position, is not unambiguous. We choose a specific 
continuation leading to 
proper analytic properties of the polarization operator, see below.)
 The inverse propagator of the $n$-th bound state can be written as
\begin{equation}
D^{-1}_n(q^2)=-(q^2-m^2_n+\Sigma (q^2 ))
\label{sigma}
\end{equation}
where $\Sigma (q^2)$
a function of order $1/N_c$ reflecting the possibility of the 
transitions 
$a\rightarrow bc\rightarrow a$. This function is known at 
$q^2=m^2_n$,
\beq
{\rm Im}\, \Sigma 
(q^2=m^2_n)=m_n\Gamma_n=\frac{Am^2_n}{\pi^3N_c} \, .
\label{sigma1}
\eeq
Here we used Eqs. (13) and (19). Now we can write 
\beq
D^{-1}_n=Q^2(1-\frac{A}{\pi^4N_c}{\rm ln}\frac{Q^2}{\Lambda^2}) 
+m^2_n.
\label{28}
\eeq
It is easy to see that (i) at $q^2=m^2_n$ Eq. (\ref{sigma1})
is satisfied; (ii) the pole is shifted to an unphysical sheet, so that on 
the 
physical sheet there are no singularities except  the cut at 
positive
 real $q^2$. The property (i) is quite obvious. Let us comment on the 
property
(ii). 

The easiest way to demonstrate that there are no singularities 
on the physical sheet is as follows. Observe, that at 
 our level of accuracy one can write, instead of Eq. (\ref{28}), 
\beq
D^{-1}_n=(z+m^2_n)
\label{29}
\eeq
where 
\beq
z=Q^2 (Q^2/2\pi^2\mu^2)^{-\frac{A}{\pi^4N_c}}\, .
\label{30}
\eeq
Here the constant $\Lambda^2$ in Eq. (\ref{28}) is adjusted
in accordance with Eq. (\ref{tilde}) below. 

The physical sheet on the complex $Q^2$ plane (Fig. 3$a$) is mapped 
onto a sheet with a 
``defect angle" on the complex $z$ plane (Fig. 3$b$).
Going into the shaded
 area we pass to the unphysical sheets. Note that the pole of $D_n$ 
lies in the 
shaded area.

\begin{figure}[htbp] \unitlength 1mm
\begin{center}
\begin{picture}(160,160)
\put(0,85){\epsfig{file=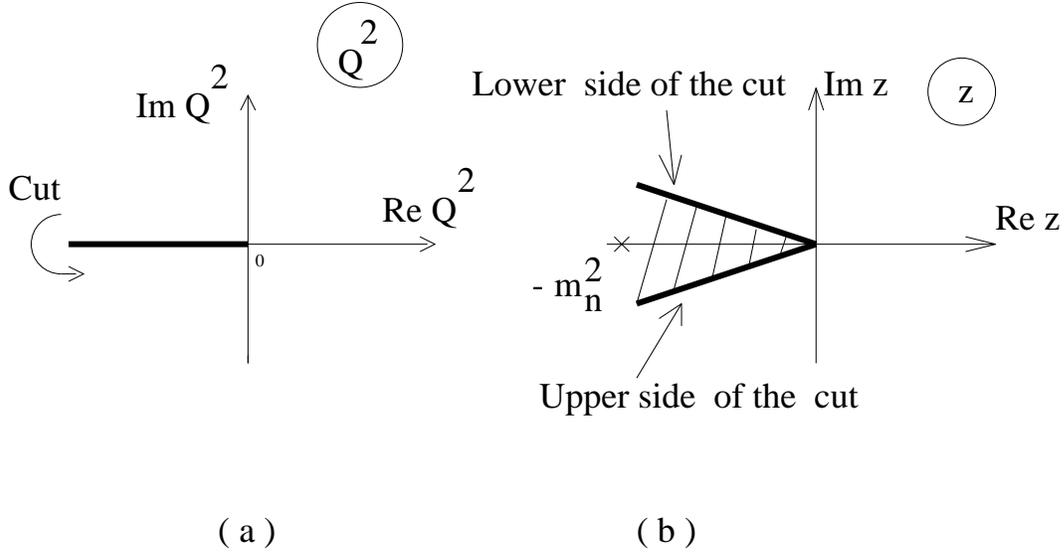,width=14cm,angle=270}}
\end{picture}
\caption{Analytical structure of the polarization operator. ($a$)
The polarization operator must be analytic everywhere in the 
complex $Q^2$ plane, except the cut running on the negative real 
semi-axis of $Q^2$ (positive real semi-axis of $q^2$). The imaginary 
part of the polarization operator must be positive at the upper side 
of the $q^2$ cut; ($b$) The mapping of the $Q^2$ plane onto the $z$ 
plane, Eq. (33). The physical sheet on the $Q^2$ plane 
corresponds to the $z$ plane with the shaded sector removed. The 
boundaries of the sector correspond to the lower and upper sides of 
the $q^2$ cut.}
\end{center}
\end{figure}

Assembling all pieces together
 we conclude that, with the resonance widths switched on, 
Eqs. (\ref{pol1}) and (\ref{psi})  are substituted by 
\beq
\Pi (Q^2)-\Pi (0) ={\rm Const.} \sum D^{-1}_n (Q^2)=-\,\, \frac{1}{1-
A/(\pi^4N_c)}\,\,
\frac{N_c}{2\pi}\,\, \psi (\tilde \sigma),
\label{31}
\eeq
where
\beq
\tilde \sigma =\frac{z}{2\pi^2\mu^2}+\frac{1}{2}\, .
\label{tilde}
\eeq
 The constant in front of $\psi (\tilde\sigma )$ 
 is adjusted in such a way as to leave intact the high $Q^2$
asymptotics, see Eq. (\ref{asymptotics}). It is  clearly  
subleading in $1/N_c$.
 Strictly speaking, we should have omitted it at the level 
of accuracy accepted here.

By construction, all singularities of the polarization operator 
(\ref{31}) are on the unphysical sheet.
 The discontinuity at the cut $q^2\ge 0$ will be calculated below.

\subsection{OPE and the asymptotics of the polarization operator}

The expression (32) for the polarization operator is clearly not exact,
since we have made a number of approximations. They do not affect, 
however,
the large $Q^2$ asymptotics of $\Pi (Q^2)$. Neither the leading  
asymptotics of high orders in the $1/Q^2$ expansion depends 
on these approximations. Since we are interested in 
the high 
energy behavior, distortions introduced in $\Pi (Q^2)$ at finite $Q^2$ 
by the
approximations made are not important.
In particular, we will disregard the fact that
low-order terms in the $1/Q^2$ expansion of $\Pi (Q^2)$
(``condensates") may come out with ``wrong" coefficients.
By adjusting the positions and residues of a few lowest resonances 
we can always get any desirable coefficients for any given
{\em finite} number of terms in the $1/Q^2$ expansion.

 Let us first study the impact of the resonance widths on 
$\Pi (Q^2)$ in the Euclidean domain. As we saw,  the effect due to 
the nonvanishing widths 
essentially reduces to the substitution of the variable $\sigma$ 
defined in 
Eq. (26) by $\tilde\sigma$, see Eq. (33). Therefore, the change in the 
asymptotic $1/Q^2$ expansion at large (Euclidean) values of $Q^2$ is 
rather 
insignificant. If at $N_c=\infty$ the $n$-th term of the power 
expansion is
 $C_n(Q^2)^{-k_n}$, at the $1/N_c$ level it becomes
\beq
C_n\frac{1}{(Q^2)^{k_n(1-\alpha)}}\rightarrow 
C_n\frac{1}{(Q^2)^{k_n}}
(1+\alpha k_n{\rm ln} Q^2)\, ,
\label{NC}
\eeq
\beq
\alpha =\frac{A}{\pi^4N_c}\, .
\eeq
The second term on the right-hand side in Eq. (\ref{NC}) is a small 
$1/N_c$ correction which 
reminds a 
logarithmic anomalous dimension in QCD. Unlike the first term in Eq.
(\ref{NC}), the logarithmic term develops an imaginary part at  large 
positive real 
values of $q^2$. If we treated Eq. (\ref{NC}) 
 in the framework of OPE (more exactly, in practical version
\cite{NSVZ}), then we would predict that the spectral density
\beq
{\rm Im} \Pi (s)\vert_{s\gg\mu^2}=\frac{N_c}{2}\left[ 1-
\sum_{n=1}^{n=n_0}C_n
\frac{\alpha\pi k_n}{(-s)^{k_n}}\right] \, ,
\label{35}
\eeq
where $s=q^2$ and $n_0$ is the highest term retained in practical 
OPE.
By construction, the prediction for the spectral density obtained from 
practical
OPE is smooth.
As a matter of fact, all correction terms in Eq. (\ref{35}) are 
suppressed
by $1/N_c$, since $\alpha\sim 1/N_c$, and are numerically 
insignificant
at large $N_c$.

Let us examine now the ``physical spectral density", i.e. the 
imaginary part  at positive $q^2$ following directly from
Eq. (\ref{31}). Our task is to reveal an
 oscillating component, not 
suppressed by $1/N_c$.

In order to find the
imaginary part analytically, it is convenient to use  the reflection 
property of the $\psi$ function,
\beq
\psi (\tilde\sigma)= \psi (-\tilde\sigma )-\pi{\rm cot} (\pi 
\tilde\sigma )
-1/\tilde\sigma\, .
\label{ref}
\eeq
It is obvious  that the polarization operator
(\ref{31}) is a sum of three terms,
corresponding to three different terms on the right-hand side
 of Eq. (\ref{ref}).
Moreover, it is evident that the first and the third terms are smooth 
functions of $q^2$.
Their contribution to the imaginary part of the polarization 
operator
for large $s$ corresponds to the smooth component obtained from 
OPE, 
see the correction terms in Eq. (\ref{35}).

 In order to study the oscillating/exponential  component
one must consider the second term,
\beq
{\rm Im}\, \Pi (s) =\frac{N_c}{2}{\rm Im}\,\,{\rm cot} 
(\pi\tilde\sigma 
)=
-\frac{N_c}{2} \frac{{\rm sinh} (2y)}{{\rm cosh} (2y)-{\rm cos} (2x)}
\, ,
\label{fet}
\eeq
where 
$$
x= \pi\, {\rm Re}\, \tilde\sigma\approx -\frac{s}{2\pi\mu^2}(1-
\alpha
 {\rm ln}(s))\, ,
$$
\beq
y=\pi\, {\rm Im}\, \tilde\sigma\approx -\frac{\alpha s}{2\mu^2}\, .
\label{xy}
\eeq
Taking into  account only the leading in $N_c$ terms it is easy
 to rewrite the 
spectral density (\ref{fet}) as follows
\beq
{\rm Im} \, \Pi (s) =
\frac{N_c}{2} \, \frac{{\rm sinh}
 (\frac{\alpha s}{\mu^2})}{{\rm cosh}
 (\frac{\alpha s}{\mu^2})-{\rm cos} (\frac{s}{\pi\mu^2})}\, .
\label{fat}
\eeq
At $\alpha s\ll \mu^2$ we are in the resonance zone (here and below 
we use
the
 nomenclature
of Ref. [3], see Sect. 5.2). In order to find the high energy behavior 
of the 
imaginary part 
of the polarization operator, we must go to the  oscillation
 zone,
to  energies $\alpha s\gg \mu^2$. Then
\beq
{\rm Im} \, \Pi (q^2) \rightarrow \frac{N_c}{2}[1+2\exp{(-
\frac{\alpha 
s}{\mu^2})}{\rm cos}
 (\frac{s}{\pi\mu^2} )]\, .
\label{exp}
\eeq
The unit term corresponds to the leading asymptotics
(it reproduces the OPE prediction), while the second term
 is an oscillating/exponential component, missing in  practical OPE
\footnote{In terms of an index $\sigma$ introduced in Sect. 5.2
of Ref. \cite{Chib1}, Eq. (\ref{exp}) implies that $\sigma = 2$.}. 
It presents  a deviation from duality we are hunting for (remember, 
by 
duality we
 understand 
a specific procedure formulated in Refs. \cite{Shif1,Chib1}).

The oscillating component is suppressed by the damping exponential,
exp $(-\alpha s/\mu^2)$. A comment is in order regarding the 
exponent,
\beq
\frac{\alpha s}{\mu^2}\equiv \frac{A}{\pi^4N_c}\frac{s}{\mu^2}\, .
\label{exp1}
\eeq
This exponent determines the boundary
between the oscillation and the
 resonance zones,
$s_0\sim \pi^4 N_c\mu^2$. This estimate is in accord with intuition.
 Indeed, the 
exponential suppression of the oscillations should start when the 
resonance
 width becomes larger than the distance between the neighboring 
resonances.
Since $\Gamma_n\sim \frac{\mu\sqrt{n}}{N_c\pi^3}$ and 
the distance between the neighboring resonances
 $\Delta m_n\sim \pi\mu/\sqrt{n}$, this occurs at the excitations 
number
$n\sim \pi^4 N_c$. Thus, the factor $1/N_c$ in Eq. (\ref{exp1}) is 
obvious.
What is more remarkable, is  a large additional numerical 
suppression,
$\pi^4$, pushing the
 boundary to higher energies and making the exponential 
suppression weaker. 
It can be traced back to the numerical suppression of the width
in Eq. (\ref{widths}) which, in turn, is due  to a strong numerical 
suppression of the phase space. Because of this fact the damping of 
the oscillations occurs very slowly in the 't Hooft model.  Figure 4
shows Im $\Pi$ in the 
window from the 
35-th to 65-th  oscillation. We see that even here the oscillation 
amplitude is 
quite sizable. 
Unlike
 other features, expected to be generic,  this numerical aspect 
-- a very slow rate of the oscillation damping -- is
specific to the 't Hooft model and is not expected to 
survive in four-dimensional  QCD, since it is entirely due to 
 ``abnormally" narrow
widths of the resonances in two-dimensional  QCD.
 Semiquantitative estimates of the damping exponent in
four-dimensional  QCD will be given in Sect. 4.

\begin{figure}
\epsfxsize=9cm
\centerline{\epsfbox{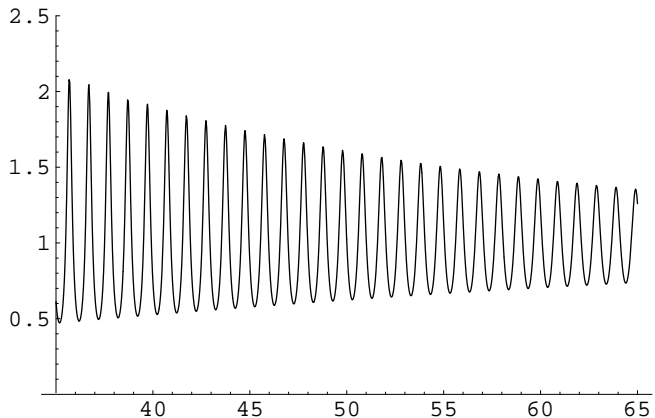}}
\caption{The spectral density corresponding to Eq. (34)
in the 't Hooft model, versus $s$ (in the units
$2\pi^2\mu^2$). $N_c$ is put equal to 3, and the normalization 
factor is chosen in such a way that asymptotically the spectral 
density dsiplayed must approach unity.}
\end{figure}

In summary, for   
$q^2
\le 0$ the polarization operator is  given by the $\psi $ 
function 
of the positive real argument and is a smooth function with no 
singularities. It is well approximated by its asymptotic expansion.
The latter is most conveniently written in terms of the variable 
$\tilde\sigma$
defined above (see Eq. (\ref{tilde})), for which it becomes  the 
standard asymptotic expansion 
of the $\psi$ function. In the Minkowskean domain,
starting from the scale $s_0\sim \pi^4\mu^2 N_c$, one can 
approximate the
 polarization
operator by its smooth asymptotics (analytically continued from the
 expansion in  the
Euclidean domain) {\it plus} an exponentially decreasing and 
oscillating term
 (\ref{exp}).

\section{Four-Dimensional  QCD}

Unlike the `t Hooft model, we cannot solve four-dimensional  QCD. At 
the qualitative 
level, however, we can apply the same ideas. The general pattern of 
the
 asymptotic
behavior will be the same, but the numerical aspects look different.

Our goal in this section is to discuss the polarization 
operator of two vector currents:
\beq
\Pi_{\mu\nu} (q^2)
 =i\int\exp{(iqx)}\langle 0\vert T\{j_\mu (x) j_\nu 
(0)\}\vert0\rangle
d^4x\, .
\label{pol4}
\eeq
Here $j_\mu (x)$ is the vector current,
\beq
j_\mu (x)=\bar u\gamma_\mu d(x)\, .
\label{current}
\eeq
Due to the conservation of the vector current  the polarization 
operator 
can be 
represented as 
\beq
\Pi_{\mu\nu} (q^2)= (q^\mu q^\nu -q^2g_{\mu\nu})\Pi (q^2)\, .
\label{polarisation}
\eeq
Furthermore, $\Pi (q^2)$ is dimensionless, and is perfectly analogous 
to 
the 
polarization operator (21) we dealt with in the 't Hooft model. 
According 
to the standard wisdom of  multicolor QCD, we expect that at 
$N_c=\infty$
and high energies $\Pi (q^2)$ is representable as a sum of 
equidistant
 infinitely narrow resonances, with a constant residue, i.e. we arrive
at the same $\psi$ function. The dynamical smearing is again 
provided by the 
resonance
widths (a $1/N_c$ effect). What is known about the widths of the 
highly excited states in four-dimensional  QCD?

In the case at hand we do not have in our disposal quantitative tools 
which would allow us to calculate the widths.  Such a calculation 
could have been performed in a non-critical string theory, were this 
theory available. Unfortunately, the issue is not worked out
\footnote{Some attempts in this direction were reported in Ref.
\cite{att}.},
and we have to resort to qualitative arguments.  If the string-based 
picture of color confinement is indeed valid, one can hardly avoid
the conclusion that the resonance widths must 
 grow linearly with 
$m_n$ \cite{Nussinov}, much in the same way as in the 't Hooft 
model.

Let us remind quasiclassical arguments of Ref. \cite{Nussinov},
showing that $\Gamma_n\sim m_n/N_c$ in four-dimensional  QCD.
When a highly excited meson state is created by a local source, it can
 be considered, quasiclassically, as a pair of (almost free) 
ultrarelativistic
quarks; each of them with energy $m_n/2$. These quarks are 
created 
at the origin, and then fly back-to-back, creating behind them a flux 
tube
of the chromoelectric field. The length of the tube $L\sim 
m_n/\Lambda^2$
where $\Lambda^2$ is the string tension. The decay probability is 
determined, to order $1/N_c$, 
 by the probability of creating an extra quark-antiquark pair. Since 
the pair creation 
can happen anywhere inside the flux tube, it is natural to expect that
\beq
\Gamma_n\sim \frac{1}{N_c} L\Lambda^2=\frac{B}{N_c}m_n
\label{?}
\eeq
where $B$ is a dimensionless coefficient of order one. 

Let us note in passing that the $1/N_c^2$ corrections due to 
creation of two quark pairs are of order $L^2/N_c^2$ within this 
picture. Since $L\sim m_n \sim \sqrt{n}$, the expansion parameter is
$\sqrt{n}/N_c$. 

Strictly speaking, the estimate (\ref{?}) must be, rather, viewed as a 
lower bound,
since the transverse fluctuations of the flux tube can increase the 
decay probability. However, most likely,  these transverse 
fluctuations will   materialize in the form of emission of 
glueballs,
a subleading $1/N_c^2$ effect, which is not considered here. It is not 
fully clear  
what impact these fluctuations may have on the 
quark-antiquark pair production. It seems plausible that they only 
affect the numerical coefficient  in Eq. (\ref{?}), which is not 
calculated anyway.
Given all naivet\`{e} of the arguments and the estimate (\ref{?}),
at the present stage it is reasonable to accept it as a working 
hypothesis.

As was mentioned, in  real QCD we expect $B\sim 1$, see 
estimates in Ref. \cite{Nussinov} and below. 
Apart from this numerical 
difference, everything else is perfectly analogous to the consideration 
we have carried out in the 't Hooft model. Hence, we conclude that in 
four-dimensional  QCD
a viable  model of the approach to asymptotics is provided by 
the same $\psi$ 
 function
\beq
\Pi (Q^2)-\Pi (0)={\rm Const.}\, \psi (\tilde \sigma )\, ,
\label{pi}
\eeq
where in the case at hand
\beq
\tilde\sigma=\left( \frac{Q^2}{\Delta m^2}\right)^{1-
\frac{B}{N_c\pi}} + C\, .
\label{o}
\eeq
Here $C$ is a numerical constant correlated with the position of the 
lowest resonance,
$$
m_0^2 = C\Delta m^2 \left(1+  O(1/N_c)\right)\,  ,\,\,\, 
\frac{\Gamma_0}{m_0} = \frac{B}{N_c}\left(1+  O(1/N_c)\right)\, .
$$
The normalization constant 
in front of the $\psi$ function in Eq. 
(\ref{pi}) will be chosen in the form 
$$
-\, \left( 1 -\frac{B}{N_c\pi}\right)^{-1}\, ,
$$
so that the imaginary part at asymptotically high $s$
is automatically normalized to unity. 
The plot of Im $\Pi$ 
is shown on Fig. 5.  Although fine details
of this spectral density are somewhat distorted compared to the
actual  experimentally measured spectral densities 
in the $e^+e^-$ annihilation or $\tau$ decays
(e.g. the $\rho$ meson width comes out $\sim 1.5$
larger than the experimental one) the qualitative similarity of our 
model with experiment is remarkable.  To substantiate the point
we display a plot of the spectral density in the vector isoscalar
channel borrowed from Ref. \cite{ALEPH} (Fig. 6). 

\begin{figure}
\epsfxsize=9cm
\centerline{\epsfbox{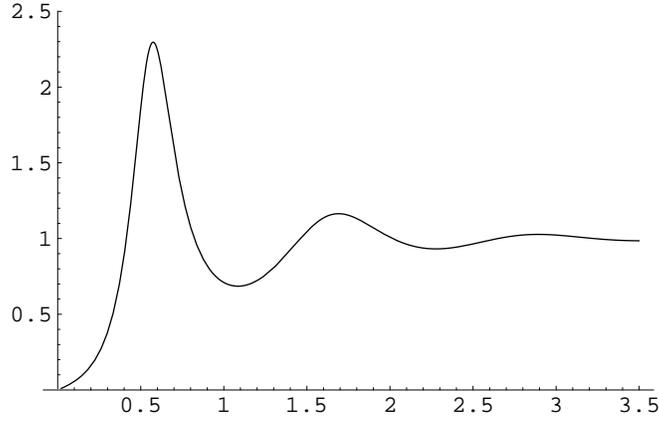}} 
\caption{The imaginary part of
the polarization operator  corresponding to Eq. (50)
with $\Delta m^2 = 1 $ GeV$^2$, $C=0.6$ and $B=0.78$. The number 
of colors is set equal to three. The horizontal axis presents $s$,
in GeV$^2$.}
\end{figure}
 
\begin{figure}
\epsfxsize=10cm
\centerline{\epsfbox{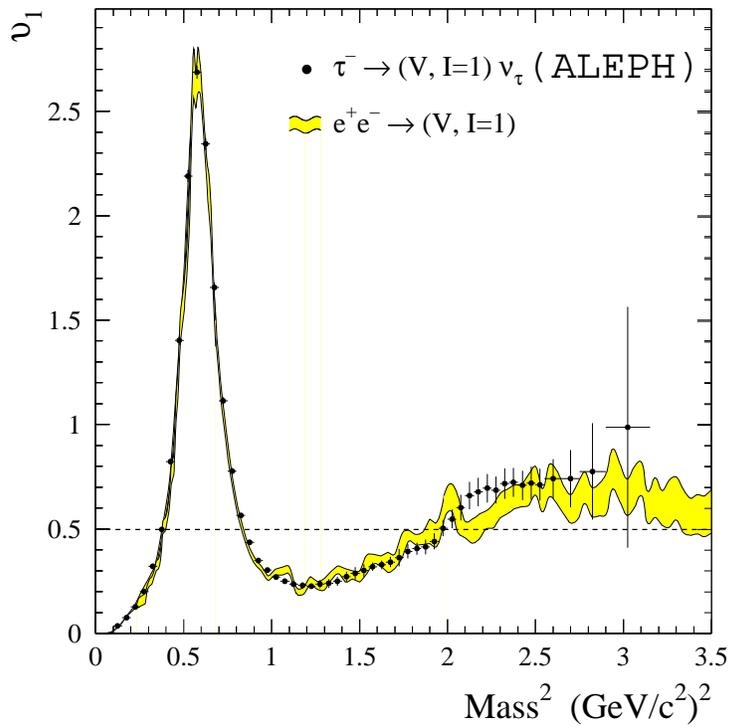}}
\caption{The spectral density in the vector isoscalar
channel measured in $e^+e^-$ annihilation and $\tau$ hadronic 
decays.}
\end{figure}

The oscillation zone starts at $s_0\sim \Delta m^2N_c(2\pi B)^{-1}$.  
Above this boundary the asymptotic form
of the oscillating exponential terms,
 analogous to  Eq. (\ref{exp}),  is given by 
\beq
{\rm Im} \Pi (Q^2)={\rm Const.}\, \left[ 1+2\exp\left( -\frac{2\pi 
Bs}{N_c\Delta 
m^2}\right) {\rm cos}
\left( \frac{2\pi s}{\Delta m^2}\right) \right]\, .
\label{tt}
\eeq
Both, the onset of the oscillation zone and the oscillation structure are
in nice qualitative agreement with what we see on Fig. 6 presenting  
experimental data on $e^+e^-$ annihilation and $\tau$ 
decays.

In summary, our model spectral density (\ref{o}),
with the appropriate values of parameters,
properly captures all important features which must be inherent to
spectral densities in real QCD. First, the corresponding polarization 
operator is a sum of an infinite number of simple poles on the 
unphysical sheet, so that the correct analytical properties ensue
automatically. The $1/Q^2$ expansion has the right structure. The 
spectral density following from Eqs. (\ref{pi}) and (\ref{o}) is 
dynamically smeared 
by the
resonance widths. Purely pictorially it  closely resembles  what is 
measured experimentally. 

\section{Conclusions}

 In this paper we address the issue of  the 
preasymptotic
 component of the spectral density not seen in practical 
 OPE.
This component oscillates, with the amplitude being exponentially 
damped. 
The approach to the  problem is complementary to  that of Refs. 
\cite{Shif1,Chib1}. It is gratifying 
to observe 
that 
the  general features of the overall picture come out the same as in 
the previous  
analyses. Details are different. For instance, the instanton-based 
model 
discussed in Refs. \cite{Shif1,Chib1} yields the interval between the 
successive 
oscillations
 growing with 
$s$, while our present result implies equidistant oscillations in the 
$s$
scale. In this aspect the instanton-based model is seemingly less 
realistic. 
Moreover, for obvious reasons it does not allow one to trace the 
proper $N_c$ dependence, while our present analysis does. 
Needless to say that it reproduces 
the desired regularity -- the fact that the exponent governing the 
exponential damping of 
the
 oscillations is proportional to $1/N_c$. 

The important lesson we draw is  confirmation of  a
general pattern of the duality-violating 
component in the 
spectral densities at high energies inferred previously:
exponential character, modulated by oscillations.
Particular details are  model-dependent. The 't Hooft model 
is 
solvable,
and all questions  can be explicitely answered. In four-dimensional  
QCD it is 
possible
to provide only educated guesses. Further efforts are needed to back 
them up 
by more solid calculations. 

We believe that the suggested  {\em ansatz} for the spectral density, 
Eqs. (\ref{pi}) and (\ref{o}),   gives a very good idea
of the duality-violating contributions. 
It is compatible with all general principles of field theory, 
$1/N_c$ expansion of QCD, and folklore knowledge which is 
universally
believed to be true. 

  Although our discussion
was phrased in terms of the spectral densities,
its implications are wider.  
In particular, it is quite probable
that  a component of the very same structure is present
in the so called  hard quantities without OPE, e.g. thrust. 
Recently it was realized that the perturbative predictions
for such quantities must be supplemented by $1/Q$ corrections
(for a review see e.g. \cite{Braun}). If we are right, in the 
intermediate domain of moderate  momentum transfers
the oscillating component might be noticeable too. 

An interesting question which deserves a new dedicated analysis
(independently of our model of duality-violating contributions)
is the behavior of the resonance widths as a function of the excitation 
number at high $n$ in multicolor four-dimensional  QCD. Another 
obvious direction 
of expansion
is combining the $\psi$ function {\em ansatz} (\ref{pi})
with information on specific low-dimension condensates
which might allow one to obtain a fully realistic description of
the polarization operator in the entire complex plane.

\vspace{0.5cm}

{\bf Acknowledgment}: \hspace{0.3cm}

We would like to  thank A. Kaidalov,  S. Nussinov and A. Vainshtein 
for  useful discussions.

This work was supported in part by DOE under the grant number
 DE-FG02-94ER40823, by the 
 Israel Academy of Sciences and the VPR Technion fund,
and by by Grant 5421-3-96
from the Ministry of Science and the Arts of Israel.

\end{document}